\documentclass[letter]{aa}     
\usepackage{graphicx}
\usepackage{txfonts}
\usepackage{hyperref}
\usepackage{multirow}
\usepackage{multicol}

\newcommand{\ipdfmp}[0]{$ipd\_frac\_multi\_peak$}

\begin{document} 

\title{
New multiple AGN systems with sub-arcsec separation : confirmation of candidates selected via the novel GMP method}
\titlerunning{Unveiling multiple AGNs via the GMP method}

\author{
A. Ciurlo\inst{1}       \and 
F. Mannucci\inst{2}     \and 
S. Yeh\inst{3}          \and 
A. Amiri\inst{2,4}      \and 
S. Carniani\inst{5}     \and 
C. Cicone\inst{6}       \and 
G. Cresci\inst{2}       \and 
R. Khatun\inst{6}      \and 
E. Lusso\inst{2,4}      \and 
A. Marasco\inst{7}      \and 
C. Marconcini\inst{2,4} \and 
A. Marconi\inst{4}      \and 
E. Nardini\inst{2}      \and 
E. Pancino\inst{2}      \and 
P. Rosati \inst{8}     \and
P. Severgnini\inst{9}   \and 
M. Scialpi\inst{4,2}    \and 
G. Tozzi\inst{4,2}      \and 
G. Venturi\inst{10,2}   \and 
C. Vignali\inst{11}     \and 
M. Volonteri\inst{12}
}

\institute{
Department of Physics and Astronomy, University of California Los Angeles, 430 Portola Plaza, Los Angeles, CA 90095, USA \email{ciurlo@astro.ucla.edu} \and 
INAF, Osservatorio Astrofisico di Arcetri, largo E. Fermi 5, 50125 Firenze, Italy \and 
W. M Keck Observatory, 65-1120 Mamalahoa Highway, Kamuela, HI 96743, USA \and 
Dipartimento di Fisica e Astronomia, Università di Firenze, Via G. Sansone 1, 50019, Sesto Fiorentino (Firenze), Italy \and 
Scuola Normale Superiore, Piazza dei Cavalieri 7, 56126, Pisa, Italy \and 
Institute of Theoretical Astrophysics, University of Oslo, P.O Box 1029, Blindern, 0315 Oslo, Norway \and 
INAF-Osservatorio Astronomico di Padova, Vicolo Osservatorio 5, Padova, Italia \and 
University of Ferrara, Department of Physics and Earth Sciences, Via G. Saragat, 21-44122 Ferrara, Italy \and
INAF, Osservatorio Astronomico di Brera, Via Brera 28,20121 Milano, Italy \and 
Instituto de Astrofísica, Facultad de Física, Pontificia Universidad Católica de Chile, Casilla 306, Santiago 22, Chile \and 
Physics and Astronomy Department "Augusto Righi", Università di Bologna, Via Gobetti 93/2, 40129 Bologna, Italy \and 
Institu d'Astrophysique de Paris,  98bis Bd Arago, 75014 Paris, France 
}   

\date{Submitted January 6, 2023}

\abstract
{
The existence of multiple active galactic nuclei (AGN) at small projected distances on the sky is due to either the presence of multiple, in-spiraling SMBHs, or to gravitational lensing of a single AGN. Both phenomena allow us to address important astrophysical and cosmological questions. 
However, few kpc-separation multiple AGN are currently known.
Recently, the newly-developed Gaia Multi peak (GMP) method provided numerous new candidate members of these populations. 
We present spatially resolved, integral-field spectroscopy of a sample of four GMP-selected multiple AGNs candidates.
In all of these systems, we detect two or more components with sub-arcsec separations. 
We find that two of the systems are dual AGNs, one is either an intrinsic triple or a lensed dual AGN, while the last system is a chance AGN/star alignment.
Our observations double the number of confirmed multiple AGNs at projected separations below 7~kpc at $z>0.5$, present the first detection of a possible triple AGN in a single galaxy at $z>0.5$, and 
successfully test the GMP method as a novel technique to discover previously unknown multiple AGNs.
}


\maketitle

\section{Introduction}
\label{sec:introduction}
All current cosmological models describe galaxy formation as a hierarchical process in which small galaxies merge to form larger systems. 
This process also applies to the supermassive black-holes (SMBHs) that co-evolve with the host galaxy \citep{begelman80}. 
Given the long merging timescale \citep[$\sim$1~Gyr, e.g][]{Tremmel17}, a population of dual or multiple SMBHs must exist in many galaxies \citep{Volonteri03}. 
SMBHs are expected to accrete material from the merging host galaxies, producing dual or multiple luminous active galactic nuclei (AGNs) 
in the same galaxy \citep{Steinborn16, Rosas-Guevara19, Volonteri22}. 
For example, \cite{Volonteri22} estimate that at $z>2$ more than 1$\%$ of the bright AGNs (L$_{\rm bol}>$10$^{43}$~erg/s) are expected to have a companion within 10~kpc.
The discovery of dual AGNs at kiloparsec-scale separation is therefore crucial to support the hierarchical formation model. 
Additionally, since dual AGNs are the precursors of a binary phase, they allow us to study the merging steps leading to 
the emission of gravitational waves \citep[e.g.][]{Colpi14}.

Several tens of dual AGNs at separations above  10--20~kpc are known \citep[e.g.][among many others]{Lemon19, Chen22}. 
However, very few dual-AGN at separations below $\sim$5~kpc --compatible with being in the same host galaxy-- have been discovered so far. 
There is a shortage of known close systems especially at intermediate and high redshifts, when galaxy mergers are more common
(see \citealt{Derosa19} and \citealt{mannucci22} and references therein).
This lack is  due to the relatively low efficiency  of the current selection techniques for sub-arcsec separations systems \citep{Rubinur19}.
In particular, only four systems with separations below 5~kpc have been confirmed at z$>$0.5 
\citep[Glikman et al., in prep]{junkkarinen01, Chen22, mannucci22}.
The small number of currently known dual AGN systems prevents us from testing cosmological model predictions such as the fraction of dual systems over the total AGN population, their evolution with redshifts and their mass and luminosity ratios \citep[][and references therein]{Volonteri22}.

Thanks to its high spatial resolution and full sky coverage, the Gaia satellite is revolutionizing the field (e.g \citealt{Lemon19,Shen21,Chen22,Lemon22}). 
In particular, the Gaia Multi-peak (GMP) method \citep{mannucci22} allows us to select large numbers of dual systems with separations down to $\sim$0.15" by searching for multiple peaks in the light profile of the Gaia sources.
\cite{mannucci22} tested the efficiency of this method on 31 GMP-selected systems with HST (archival images of 26 systems) and LBT (newly obtained high-resolution observations of five systems) images. 
All these systems show multiple compact sources with sub-arcsec resolution, confirming that this novel technique can be extremely efficient in selecting a sample of quasi-stellar objects with multiple components. 

The GMP-identified systems can also be
lensed, high-redshift AGN, that appear as multiple components with small spatial separations. 
Strongly lensed AGN are rare and unique tools for measuring the Hubble parameter (e.g. \citealt{wong18}) and for investigating AGN feedback at high redshift (e.g. \citealt{feruglio17, tozzi21}). 
In particular, very compact systems (sub-arcsec separations) allow us to investigate the mass distribution of lensing galaxies to a regime lower than what is typically probed by current galaxy-scale lenses surveys
\citep[e.g., SLACS][]{Bolton08, Shajib21}. 
The sensitivity to such low-mass dark matter halos can be used to study the nature of dark matter \citep[e.g.][]{casadio21}.

A crucial next step is to understand the nature of the GMP-selected systems: intrinsically multiple AGNs, gravitationally-lensed systems or an AGN plus a foreground star. 
Integral field spectroscopy is particularly well-suited to extract spatially-resolved spectra of each component of these systems, thus helping us discriminate among these three scenarios.
Here, we present the first spatially resolved spectroscopy of four GMP-selected systems, observed with the adaptive optics (AO) integral field spectrograph OSIRIS at W. M. Keck Observatory \citep{Larkin06}.
The goals of these observations are: 
(1) resolving point-sources in dual-AGN candidates 
to test the success rate of the GMP technique; 
(2) differentiating AGNs from stars in resolved systems, based on  their spectral properties; 
(3) classifying the systems as intrinsically multiple vs. lensed AGNs, based on the differences between their spectra.

This letter is structured as follows. Observations and data reduction are reported in Section~\ref{sec:observations}, the classification of each system is discussed in  Section~\ref{sec:results}. Our conclusions are summarized in Section~\ref{sec:discussion}. 
All magnitudes we report are in Vega system and we used the cosmological parameters from \cite{Planck2018}.

\begin{figure*}[t]
    \centering
    \includegraphics[width=0.99\textwidth]{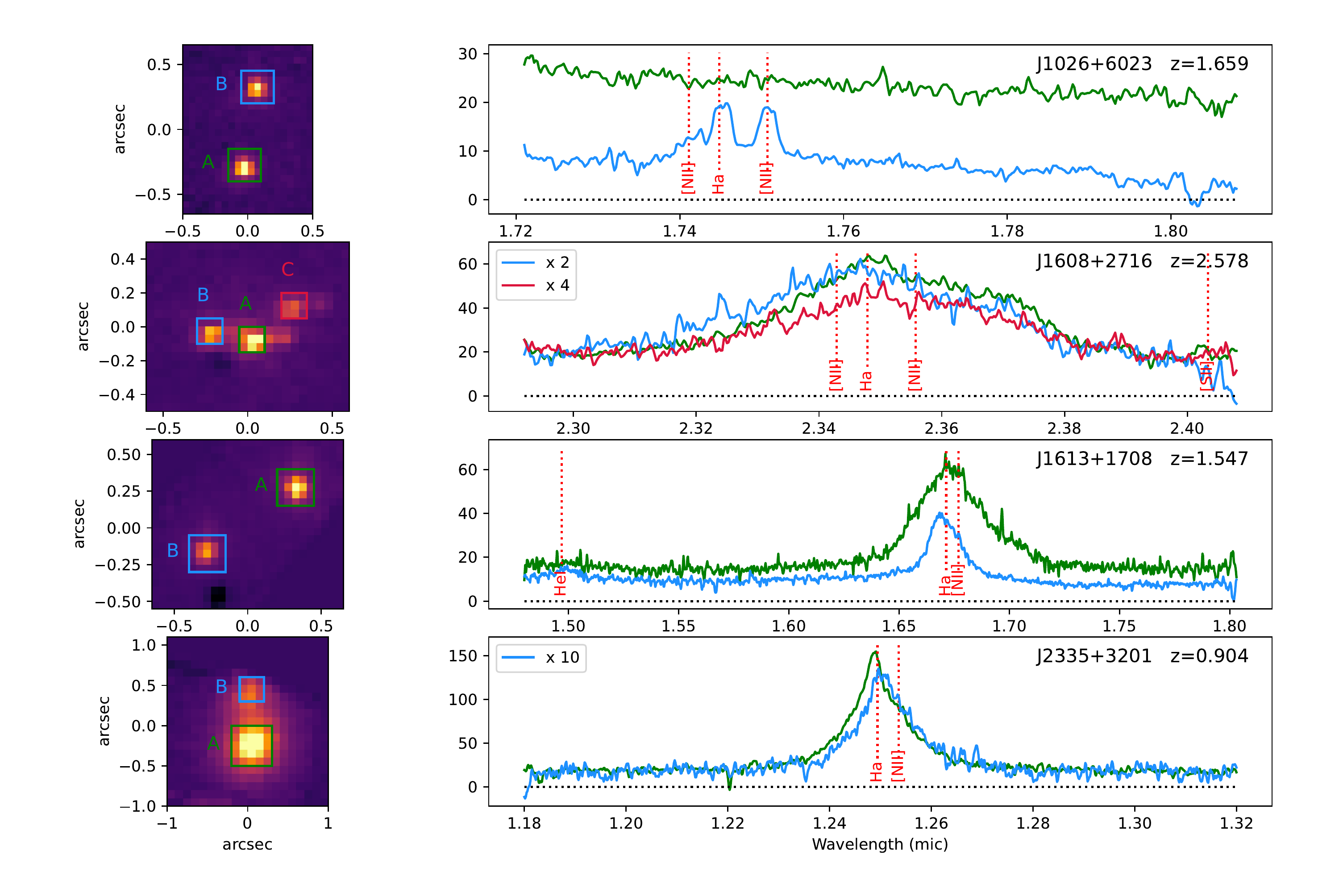}
    \caption{H$\alpha$ emission line maps (left) and spectra (right) of the systems observed with OSIRIS (target name and redshift reported in the right panels). The line maps are oriented with North up and West right. The spectra shown on the right panels have been extracted over the squared apertures marked on the left panes (with the same color-coding) . Each component of the systems is labelled as in Table~\ref{tab:results}. To optimize the visualization some of the spectra have been multiplied by the factors indicated in the labels. Vertical dotted lines show the position of the main expected emission lines.
    }
    \label{fig:spectra}
\end{figure*}

\begin{table*}[]
\begin{center}
\begin{tabular}{lccccclcccc}
\hline
\hline
Target     & RA         & DEC      & PA   & IPDfmp& Redshift
& Band   & T$_{exp}\times$N$_{exp}$ & FWHM  & seeing & AO \\
\hline
\object{J1026+6023} & 10:26:31.13& +60:23:30.13&  102$^\circ$ & 21 & 1.660  & Hn5  & 900~s $\times$4 & 0.10''& 0.7''& LGS \\
\object{J1608+2716} & 16:08:29.23& +27:16:26.74& -357$^\circ$ & 14 & 2.575  & Kn5  & 900~s $\times$6 & 0.09''& 0.7''& LGS \\
\object{J1613+1708} & 16:13:20.01& +17:08:39.40&  135$^\circ$ & 14 & 1.547  & Hbb  & 900~s $\times$4 & 0.11''& 0.7''& LGS \\
\object{J2335+3201} & 23:35:22.52& +32:01:09.08& -106$^\circ$ & 13 & 0.904  & Jbb  & 600~s $\times$2 & 0.42''& 0.9''& NGS \\
\hline
\end{tabular}
\end{center}
\caption{
Main properties of the four targets studied in this work, along with Keck OSIRIS observational setup.
IPDfmp is the value of the \ipdfmp\ parameter of the Gaia archive used for the GMP selection. 
Redshift are obtained from SDSS ground-based spectra, as reported in the Milliquas catalog.
FWHMs are calculated on isolated sources. 
The seeing corresponds to the DIMM (Differential Image Motion Monitor) seeing mean value (at zenith, at 0.5~$\mu$m), as reported by the \href{http://mkwc.ifa.hawaii.edu/current/seeing/}{Maunakea Weather Center} 
for the same night of the observations.
}
\label{tab:observations}
\end{table*}

\section{Target selection, observations, and data reduction}
\label{sec:observations}

Our targets were extracted from the Milliquas v7.2 catalog \citep{flesch21} by selecting systems observable from Keck, with spectroscopic redshifts $z>0.5$, and redshift such as to have at least one bright line (H$\alpha$ for all these systems) inside one of near-IR bands used by OSIRIS. 
All sources were selected through the GMP method by having values of \ipdfmp \footnote{the parameter of the Gaia archive used for the GMP selection} above the threshold of 10 \citep{mannucci22}. 
We exclude objects where clear stellar features at zero velocity in their archival ground-based spectrum reveals the presence of a chance alignment between an AGN and a foreground star.

All observations and observing conditions are reported in Table~\ref{tab:observations}.
We observed systems \object{J1026+6023}, \object{J1608+2716} and \object{J1613+1708} on March 19th 2022 with laser guide star (LGS) AO, with a 50~mas pixelscale. 
On our second scheduled observing date, August 12th 2022, the laser was not available, so we observed system \object{J2335+3201} with a natural guide star (NGS) correction instead. The tip and tilt star for this target is faint (14.33 magnitudes in R band, fainter than Keck’s nominal NGS limit), therefore the correction was worse than during our other observations.
Given the lower spatial resolution provided by this correction, we opted for a larger pixelscale of 100~mas.

Due to their relatively large separation (0.75" and 0.61", respectively), systems \object{J1613+1708} and \object{J2335+3201} are already resolved into two sources in the Gaia archive. 
This allows us to know the separation angle and the system orientation in advance. 
Therefore, we used the small OSIRIS field of view (0.8"$\times$3.2" at 50~mas platescale, 1.6"$\times$6.4" at 100 mas platescale) which corresponds to broad-band filters
(respectively Hbb from 1.473 to 1.803~$\mu$m and Jbb from 1.180 to 1.416~$\mu$m). 
The other two targets (J1026+6023 and J1608+1716) appear as single entries in the Gaia archive. 
Therefore, we observed them with a larger field of view (1.6"$\times$3.2") that allowed us to account for the unknown orientation of the systems but that comes with a narrower spectral coverage (respectively Hn5 from 1.721 to 1.808~$\mu$m and Kn5 from 2.292 to 2.408~$\mu$m).
In addition to the science targets, each night we also observed a standard star of spectral type A for telluric calibration, and a field of view free of targets for sky subtraction.
All data cubes were assembled and reduced using the standard OSIRIS pipeline \citep{Lockhart19}. 

For each target, we extract the spectrum of all detected components by taking the weighted sum in the squared apertures shown in the Figure~\ref{fig:spectra} (left panels). 
We calculate the weighting factor for each spaxel by extracting its corresponding spectrum and measuring the total H$\alpha$ flux. 
In this way, the signal-to-noise is maximized while the cross-contamination between different components and the aperture size impact are minimized.
We note that this technique applies because the sources are expected to be point-like and, therefore, to show no spectral variation across the field of view.

\begin{table*}[]
\centering
\begin{tabular}{clcclcc}
\hline
\hline
Target       &  Class     & \multicolumn{2}{c}{Separation} & Line & Center & redshift  \\
             &            & arcsec                         & kpc  &       & ($\mu$m) & \\
\hline
\object{J1026+6023A}  & AGN        &         &     & H$\alpha$       & 1.7451 &  1.659\\
                      &            &         &     & [NII]6854       & 1.7503 &  1.667\\
\object{J1026+6023B}  & Star       &  0.61   & -   & -               & -      &   -   \\
\hline
\object{J1608+2716A}  & \multirow{3}{*}{dual/triple AGN} &         &     & H$\alpha$+[NII] & 2.3524 & 2.584\\
\object{J1608+2716B}  &  & 0.25    & 2.0 & H$\alpha$+[NII] & 2.3467 &  2.576\\
\object{J1608+2716C}  &  & 0.29    & 2.4 & H$\alpha$+[NII] & 2.3527 &  2.585\\
\hline
\object{J1613+1708A}  & \multirow{2}{*}{dual AGN}   &         &     & H$\alpha$+[NII] & 1.6732 &  1.550\\
\object{J1613+1708B}  &    & 0.71    & 6.1 & H$\alpha$+[NII] & 1.6702 &  1.545\\
\hline
\object{J2335+3201A}  & \multirow{2}{*}{dual AGN}   &         &     & H$\alpha$+[NII] & 1.2492 &  0.904\\
\object{J2335+3201B}  &    & 0.61    & 4.8 & H$\alpha$+[NII] & 1.2508 &  0.906\\
\hline
\end{tabular}
\caption{Summary of the results from our OSIRIS observations: most probable classification, projected angular and linear distances from the brightest object, and center of the observed lines.}
\label{tab:results}
\end{table*}

\section{Results}
\label{sec:results}

We find that all four targets are resolved into multiple point-sources, with separations in the expected range \citep{mannucci22}. The images and the spectra of all the systems are shown in Figure~\ref{fig:spectra}. 
These spatially-resolved spectra allow us to study the nature of each object, as summarized in Table~\ref{tab:results}.

\subsection{J1026+6023}

\object{J1026+6023} is composed of an AGN and a star. 
The AGN shows a H$\alpha$ line with broad and narrow components and a prominent narrow [NII]$\lambda$6584 line with a redshift of z=1.659. 
This AGN is at 0.61" separation from an object with a featureless spectrum which we identify as a foreground star. 
The AGN (component A) is the brightest object in the optical band, sampled by Gaia and the Sloan Digital Sky Survey (SDSS, \citealt{Lyke20}), while the star is the brightest object in the near-IR H band sampled by the Keck spectra (component B). 
Chance AGN/star alignments of this kind are expected to be 30$\%$ of the GMP-selected targets \citep{mannucci22}.

\subsection{J1608+2716}

\object{J1608+2716} is an obscured quasi-stellar object (QSO), at z=2.575, with $A_{V}\sim$1.8 as estimated from the SDSS spectrum. 
Our observations reveal three components: a central brightest one (component A),  one 0.25" to the east (component B) and one 0.29" towards north west (component C).
Faint extensions are visible for components A and C, but their low luminosity, compared with nearby components, and the extended wings of the AO point-spread-function (PSF) do not allow us to extract independent spectra. 
Due to the shorter wavelength range used in the observations, the spectra only cover the broad H$\alpha$ line and a limited part of the continuum on both sides. 
All the three components show broad H$\alpha$ lines at similar redshifts,  with velocity dispersion of about 5500~km/sec full width at half maximum (FWHM), 
but with slightly offset line centers. 

There are three main possible explanations to a triple object:  
1) a triple lensed system, i.e, three images of the same object; 
2) lensing of a dual AGN: two distinct objects, one of which with two detected lensed images;
3) a systems of three different AGNs, a possibility predicted by current models (e.g \citealt{Ni22, Bhowmick20, Volonteri22}) and previously observed in the Local Universe (e.g \citealt{Foord21, Yadav21}). 

\begin{figure*}
  \centering
    \includegraphics[width=0.99\textwidth]{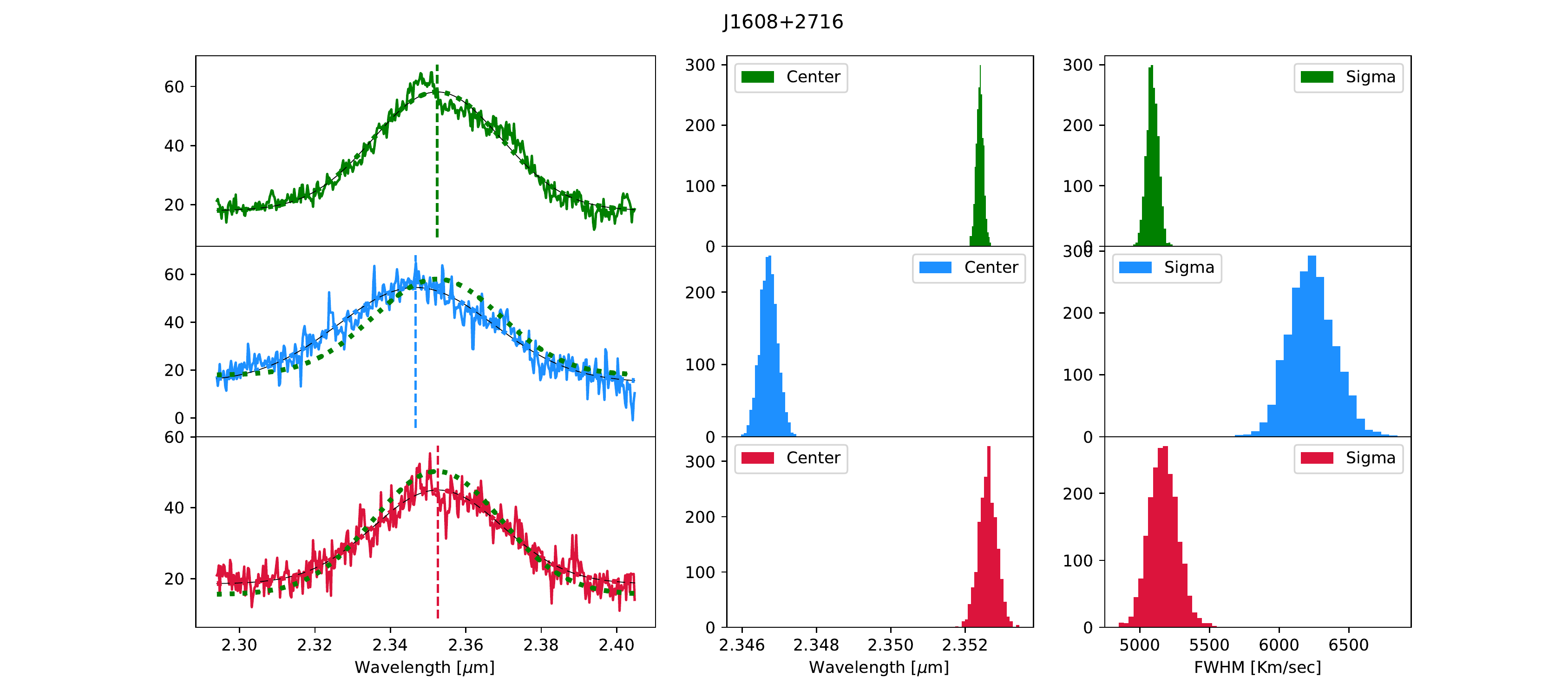}
    \caption{Comparison of the H$\alpha$ lines of the three components of J1608+2716. From top to bottom: A, B and C components, color-coded as in Figure~\ref{fig:spectra}. 
    Left panels: observed emission line (solid, thick line) and fit with a Gaussian profile plus a constant (think solid line). The center of the best-fitting Gaussian is reported as a vertical dashed line. 
    In all panels the green dotted line show the fit to A (the brightest component) for comparison.
    Center and right columns: centroids (center) and FWHM (right) distributions determined by our Gaussian fit on 2000 stochastic realisations of the observed spectra, each obtained by injecting noise into the data.
    }
    \label{fig:J1608}
\end{figure*}

To unveil the nature of this source we can consider the following points: 
\begin{itemize}

\item{Line position and profile:}
component B displays both a different line profile and radial velocity with respect to the central, brightest component A, as shown in Figure~\ref{fig:J1608}.
Gaussian fits to the emission lines of all components show that the H$\alpha$ line of component B (in blue) is centered at lower wavelengths, with a difference of $\sim$720~km/sec, and has a FWHM larger than component A by 1200~km/sec. 
We estimated the uncertainties on the center and the FWHM of the best-fit Gaussians by adding Gaussian noise to the spectra at the observed amplitude, and computing the fit again. This process was repeated 2000 times for each line. 
The distribution of the resulting centers and FWHM are show in Figure~\ref{fig:J1608} (center and left panels). 
This shows that the differences in center and width between components A and B are highly significant. 
We can exclude spatially-dependent calibration issues because the sky lines in spectra extracted at the locations of the components overlap perfectly. 
In contrast to component B, component C has a spectrum compatible with A.

\item{Variability and time lag:} 
given the small projected separation (0.25"), in the case of lensing, the time delay between components A and B is 2 days at most \citep{lieu08}. 
For intrinsic variability to be at the origin of the differences above, this timescale must be larger than (or of the same order of) the size of the broad-line emitting region (BLR).
\cite{Bentz13} have estimated the radius of the Balmer-line emitting part of the BLR as a function of the luminosity of the continuum $\lambda L(\lambda)$ at 5100~\AA. 
For J1608+2716, this luminosity --estimated from the SDSS spectrum and the G-band Gaia magnitude-- is log($\lambda~L(\lambda$))/erg~sec$^{-1}$=46.0$\pm$0.2. 
For this luminosity, \cite{Bentz13} estimate a radius of the BLR of $\sim400$~light-days. 
Even assuming that the luminosity of this object is boosted by a factor of 10 by lensing, the radius would be $\sim100$~light-days. 
This is much larger than the expected delay. 
Therefore, in the case of lensing, no significant variability of the H$\alpha$ line would be expected between the two images.

\item{Lensing:} 
component C (red in Figure~\ref{fig:J1608}) has center and FWHM compatible with the brightest component A. However, the two lines have significantly different equivalent widths (377\AA\ for component A, vs. 232\AA\ for component C). This difference, in the lensing scenario, could be attributed to microlensing of the continuum by single stars in the lensing galaxy \citep[e.g.][]{Hutsemkers10}.
If A and C are lensed images of the same QSO, the B image would be the second component of a dual AGN, however producing a single image if it lies outside the radial caustic of a general elliptical mass distribution. 
In any case, a compact lensing galaxy should be present.

\item{Missing lensing galaxy:}
nothing is detected in the observed spectra besides the QSOs and the faint extensions of component A and C.
Two lensed images of a QSO at $z_s=2.57$ separated by 0.25" (with a third image strongly demagnified near the center) can be obtained with a lens galaxy with a mass of
 M$\sim 10^{10}$M$_\odot$, by assuming it at redshift $z_L\sim 0.5-1$ and by requiring the separation to be twice the Einstein radius of a singular isothermal sphere\footnote{$\theta_E^{SIS}=[D_{LS}/(D_L D_s)\, 4GM/c^2]^{1/2}$, where $D_L$, $D_S$ are the angular diameter distances of the lens, the source and $D_{LS}$ the one between the lens and the source.}. Such a compact lensed system would only sample the central part of the lensing galaxy where the contribution of dark matter is gravitationally subdominant with respect to stellar mass, with a contribution lower than the uncertainties. Assuming that this mass is dominated by stars, we estimate a galaxy magnitude between Ks$\sim$19.2 at z=0.5 and Ks$\sim$20.5 at z=1.0 \citep[][for an early-type galaxy with a Chabrier initial mass function]{Longhetti09}. As a comparison, the QSO has Ks$\sim$19.1, estimated using Gaia magnitudes and SDSS spectra. A lensed galaxy at z=0.5 would, therefore, be easily detected also considering that it is not a point source, while would be below detection at z=1, especially if it dust extincted. The nucleus of the lensing galaxy could be the faint extension of component A, that otherwise could be the QSO host galaxy.
\end{itemize}

In conclusion, the differences in line center and profile between components A and B, together with the small time delay between the  images, show that this is not a single, triply-imaged lensed QSO, but that at least two components must be present. Components A and C are compatible with a double lens system with some contribution from microlensing, with the possible detection of the host galaxy. This system would be a lensed dual QSO, similar to the system described by \cite{Lemon22}. 
However, since a foreground lensing galaxy is not clearly detected, this system could also be a physically triple AGN. Some knowledge of the spectral energy distribution of the three sources would further help to understand the nature of this system.  

\subsection{J1613+1708} 

J1613+1708 is a very blue QSO, with no evidence for dust extinction in the SDSS spectrum. 
We find that this system shows two components with similar luminosities and a separation of 0.71" (6.1 kpc). 
A bright H$\alpha$ line is present in both spectra, with a velocity shift of $\sim500$~km/sec, corresponding to redshifts of z=1.550 and z=1.545 respectively. 
The line width are also very different: 6200~km/sec FWHM for component A, and 3100~km/sec for component B. 
In case of lensing, given its luminosity at 5100~\AA\ of
log($\lambda~L(\lambda))=45.2\pm0.1$, no significant variations of the H$\alpha$ line are expected on timescales shorter than 160~days (or 50~days assuming a lensing magnification by a factor of 10 \citealt{Bentz13}). 
In contrast, the delay expected due to the separation of the two components would be 10~days at most.
As a consequence, we conclude that the two objects are associated with two different AGNs in a single host.

\subsection{J2335+3201} 
This is a low-extinction ($A_V$$\sim$0.4, estimated from the SDSS spectrum) system at z$\sim$0.9 showing two distinct components 0.61" (4.8~kpc) away, with a large ($\sim12$) luminosity ratio. 
We find that both objects show a broad H$\alpha$ line width (FWHM=2900~km/sec for the component A and 2700~km/sec for component B). 
The two lines show a significant velocity shift of about 400~km/sec, and different line profiles. 
The system has log($\lambda L_\lambda$)=44.9 at 5100~\AA, implying variability timescales of $\sim$100~days (30~days in case of a lensing magnification by a factor of 10), to be compared with the expected delay of 2~days. 
Therefore, also in this case, the differences are better explained by a dual AGN system.

\section{Conclusions}
\label{sec:discussion}

We used AO-assisted, spatially-resolved spectroscopy to unveil the nature of four complex AGN systems at redshifts between 0.9 and 2.4 selected through the GMP method.  
As expected by the GMP selection, all these objects show multiple components with sub-arcsec separations. Target J1026+6023 is better described by a AGN/star alignment (given the featureless continuum), while emission from broad lines typical of QSO are seen in all the components of the remaining three systems. Velocity shifts of a few hundreds km/sec are seen in J1608+2716, J1613+1708 and J2335+3201, compatible with being due to multiple distinct SMBHs likely to be in the process of merging inside a single host. The differences in line profiles and projected separations are indeed best reproduced by intrinsically distinct SMBHs rather than lensing by a foreground galaxy. In fact, the luminosity of the three QSOs, even allowing for possible lensing magnification, implying large sizes of the BLR and therefore slow variability on timescales of several tens/hundreds of days. 
Since the expected time delay between different lensed images would correspond to a few days at most, the differences cannot be due to lensing delay. 
Moreover, there is no evidence for a foreground lensing galaxy.
These observations confirm that a sizeble sample of intrinsic multiple AGNs can be obtained with a reasonable amount of resolved spectra of GMP selected systems. Future observations from the ground (especially with VLT/MUSE, VLT/ERIS, and Keck/OSIRIS) and from the space (HST/STIS, JWST) will allow us to largely increase the number of confirmed multiple systems and begin to compare the results with theoretical predictions on galaxy formation and evolution. \\

\begin{acknowledgements} 
AC acknowledges support from NSF AAG grant AST-1412615, Jim and Lori Keir, the W. M. Keck Observatory Keck Visiting Scholar program, the Gordon and Betty Moore Foundation, the Heising-Simons Foundation, and Howard and Astrid Preston. 
GC, FM, AM and EN acknowledge support by INAF Large Grants "The metal circle: a new sharp view of the baryon cycle up to Cosmic Dawn with the latest generation IFU facilities"
and “Dual and binary supermassive black holes in the multi-messenger era: from galaxy mergers to gravitational waves”
(Bando Ricerca Fondamentale INAF 2022).
GV acknowledges support from ANID program FONDECYT Postdoctorado 3200802.
The authors wish to recognize and acknowledge the very significant cultural role and reverence that the summit of Maunakea has always had within the indigenous Hawaiian community. We are most fortunate to have the opportunity to conduct observations from this mountain.
\end{acknowledgements}

\bibliographystyle{aa} 
\bibliography{references}{}

\begin{thebibliography}{34}
\expandafter\ifx\csname natexlab\endcsname\relax\def\natexlab#1{#1}\fi

\bibitem[{Begelman {et~al.}(1980)Begelman, Blandford, \& Rees}]{begelman80}
Begelman, M.~C., Blandford, R.~D., \& Rees, M.~J. 1980, {\textbackslash}nat,
  287, 307

\bibitem[{{Bentz} {et~al.}(2013){Bentz}, {Denney}, {Grier}, {Barth},
  {Peterson}, {Vestergaard}, {Bennert}, {Canalizo}, {De Rosa}, {Filippenko},
  {Gates}, {Greene}, {Li}, {Malkan}, {Pogge}, {Stern}, {Treu}, \&
  {Woo}}]{Bentz13}
{Bentz}, M.~C., {Denney}, K.~D., {Grier}, C.~J., {et~al.} 2013, \apj, 767, 149

\bibitem[{{Bhowmick} {et~al.}(2020){Bhowmick}, {Di Matteo}, \&
  {Myers}}]{Bhowmick20}
{Bhowmick}, A.~K., {Di Matteo}, T., \& {Myers}, A.~D. 2020, \mnras, 492, 5620

\bibitem[{{Bolton} {et~al.}(2008){Bolton}, {Treu}, {Koopmans}, {Gavazzi},
  {Moustakas}, {Burles}, {Schlegel}, \& {Wayth}}]{Bolton08}
{Bolton}, A.~S., {Treu}, T., {Koopmans}, L. V.~E., {et~al.} 2008, \apj, 684,
  248

\bibitem[{{Casadio} {et~al.}(2021){Casadio}, {Blinov}, {Readhead}, {Browne},
  {Wilkinson}, {Hovatta}, {Mandarakas}, {Pavlidou}, {Tassis}, {Vedantham},
  {Zensus}, {Diamantopoulos}, {Dolapsaki}, {Gkimisi}, {Kalaitzidakis},
  {Mastorakis}, {Nikolaou}, {Ntormousi}, {Pelgrims}, \& {Psarras}}]{casadio21}
{Casadio}, C., {Blinov}, D., {Readhead}, A.~C.~S., {et~al.} 2021, \mnras, 507,
  L6

\bibitem[{{Chen} {et~al.}(2022){Chen}, {Hwang}, {Shen}, {Liu}, {Zakamska},
  {Yang}, \& {Li}}]{Chen22}
{Chen}, Y.-C., {Hwang}, H.-C., {Shen}, Y., {et~al.} 2022, \apj, 925, 162

\bibitem[{{Colpi}(2014)}]{Colpi14}
{Colpi}, M. 2014, \ssr, 183, 189

\bibitem[{{De Rosa} {et~al.}(2019){De Rosa}, {Vignali}, {Bogdanovi{\'c}},
  {Capelo}, {Charisi}, {Dotti}, {Husemann}, {Lusso}, {Mayer}, {Paragi},
  {Runnoe}, {Sesana}, {Steinborn}, {Bianchi}, {Colpi}, {del Valle}, {Frey},
  {Gab{\'a}nyi}, {Giustini}, {Guainazzi}, {Haiman}, {Herrera Ruiz},
  {Herrero-Illana}, {Iwasawa}, {Komossa}, {Lena}, {Loiseau}, {Perez-Torres},
  {Piconcelli}, \& {Volonteri}}]{Derosa19}
{De Rosa}, A., {Vignali}, C., {Bogdanovi{\'c}}, T., {et~al.} 2019, \nar, 86,
  101525

\bibitem[{{Feruglio} {et~al.}(2017){Feruglio}, {Ferrara}, {Bischetti},
  {Downes}, {Neri}, {Ceccarelli}, {Cicone}, {Fiore}, {Gallerani}, {Maiolino},
  {Menci}, {Piconcelli}, {Vietri}, {Vignali}, \& {Zappacosta}}]{feruglio17}
{Feruglio}, C., {Ferrara}, A., {Bischetti}, M., {et~al.} 2017, \aap, 608, A30

\bibitem[{Flesch(2021)}]{flesch21}
Flesch, E.~W. 2021, arXiv:2105.12985 [astro-ph], arXiv: 2105.12985

\bibitem[{{Foord} {et~al.}(2021){Foord}, {G{\"u}ltekin}, {Runnoe}, \&
  {Koss}}]{Foord21}
{Foord}, A., {G{\"u}ltekin}, K., {Runnoe}, J.~C., \& {Koss}, M.~J. 2021, \apj,
  907, 71

\bibitem[{{Hutsem{\'e}kers} {et~al.}(2010){Hutsem{\'e}kers}, {Borguet},
  {Sluse}, {Riaud}, \& {Anguita}}]{Hutsemkers10}
{Hutsem{\'e}kers}, D., {Borguet}, B., {Sluse}, D., {Riaud}, P., \& {Anguita},
  T. 2010, \aap, 519, A103

\bibitem[{{Junkkarinen} {et~al.}(2001){Junkkarinen}, {Shields}, {Beaver},
  {Burbidge}, {Cohen}, {Hamann}, \& {Lyons}}]{junkkarinen01}
{Junkkarinen}, V., {Shields}, G.~A., {Beaver}, E.~A., {et~al.} 2001, \apjl,
  549, L155

\bibitem[{{Larkin} {et~al.}(2006){Larkin}, {Barczys}, {Krabbe}, {Adkins},
  {Aliado}, {Amico}, {Brims}, {Campbell}, {Canfield}, {Gasaway}, {Honey},
  {Iserlohe}, {Johnson}, {Kress}, {LaFreniere}, {Lyke}, {Magnone}, {Magnone},
  {McElwain}, {Moon}, {Quirrenbach}, {Skulason}, {Song}, {Spencer}, {Weiss}, \&
  {Wright}}]{Larkin06}
{Larkin}, J., {Barczys}, M., {Krabbe}, A., {et~al.} 2006, in \procspie, Vol.
  6269, Society of Photo-Optical Instrumentation Engineers (SPIE) Conference
  Series, 62691A

\bibitem[{{Lemon} {et~al.}(2022){Lemon}, {Millon}, {Sluse}, {Courbin}, {Auger},
  {Chan}, {Paic}, \& {Agnello}}]{Lemon22}
{Lemon}, C., {Millon}, M., {Sluse}, D., {et~al.} 2022, \aap, 657, A113

\bibitem[{{Lemon} {et~al.}(2019){Lemon}, {Auger}, \& {McMahon}}]{Lemon19}
{Lemon}, C.~A., {Auger}, M.~W., \& {McMahon}, R.~G. 2019, \mnras, 483, 4242

\bibitem[{Lieu(2008)}]{lieu08}
Lieu, R. 2008, The Astrophysical Journal, 674, 75, aDS Bibcode:
  2008ApJ...674...75L

\bibitem[{{Lockhart} {et~al.}(2019){Lockhart}, {Do}, {Larkin}, {Boehle},
  {Campbell}, {Chappell}, {Chu}, {Ciurlo}, {Cosens}, {Fitzgerald}, {Ghez},
  {Lu}, {Lyke}, {Mieda}, {Rudy}, {Vayner}, {Walth}, \& {Wright}}]{Lockhart19}
{Lockhart}, K.~E., {Do}, T., {Larkin}, J.~E., {et~al.} 2019, \aj, 157, 75

\bibitem[{{Longhetti} \& {Saracco}(2009)}]{Longhetti09}
{Longhetti}, M. \& {Saracco}, P. 2009, \mnras, 394, 774

\bibitem[{{Lyke} {et~al.}(2020){Lyke}, {Higley}, {McLane}, {Schurhammer},
  {Myers}, {Ross}, {Dawson}, {Chabanier}, {Martini}, {Busca}, {Mas des
  Bourboux}, {Salvato}, {Streblyanska}, {Zarrouk}, {Burtin}, {Anderson},
  {Bautista}, {Bizyaev}, {Brandt}, {Brinkmann}, {Brownstein}, {Comparat},
  {Green}, {de la Macorra}, {Mu{\~n}oz Guti{\'e}rrez}, {Hou}, {Newman},
  {Palanque-Delabrouille}, {P{\^a}ris}, {Percival}, {Petitjean}, {Rich},
  {Rossi}, {Schneider}, {Smith}, {Vivek}, \& {Weaver}}]{Lyke20}
{Lyke}, B.~W., {Higley}, A.~N., {McLane}, J.~N., {et~al.} 2020, \apjs, 250, 8

\bibitem[{{Mannucci} {et~al.}(2022){Mannucci}, {Pancino}, {Belfiore}, {Cicone},
  {Ciurlo}, {Cresci}, {Lusso}, {Marasco}, {Marconi}, {Nardini}, {Pinna},
  {Severgnini}, {Saracco}, {Tozzi}, \& {Yeh}}]{mannucci22}
{Mannucci}, F., {Pancino}, E., {Belfiore}, F., {et~al.} 2022, Nature Astronomy,
  6, 1185

\bibitem[{{Ni} {et~al.}(2022){Ni}, {DiMatteo}, {Chen}, {Croft}, \&
  {Bird}}]{Ni22}
{Ni}, Y., {DiMatteo}, T., {Chen}, N., {Croft}, R., \& {Bird}, S. 2022, arXiv
  e-prints, arXiv:2209.01249

\bibitem[{{Planck Collaboration} {et~al.}(2020){Planck Collaboration},
  {Aghanim}, {Akrami}, {Ashdown}, {Aumont}, {Baccigalupi}, {Ballardini},
  {Banday}, {Barreiro}, {Bartolo}, {Basak}, {Battye}, {Benabed}, {Bernard},
  {Bersanelli}, {Bielewicz}, {Bock}, {Bond}, {Borrill}, {Bouchet}, {Boulanger},
  {Bucher}, {Burigana}, {Butler}, {Calabrese}, {Cardoso}, {Carron},
  {Challinor}, {Chiang}, {Chluba}, {Colombo}, {Combet}, {Contreras}, {Crill},
  {Cuttaia}, {de Bernardis}, {de Zotti}, {Delabrouille}, {Delouis}, {Di
  Valentino}, {Diego}, {Dor{\'e}}, {Douspis}, {Ducout}, {Dupac}, {Dusini},
  {Efstathiou}, {Elsner}, {En{\ss}lin}, {Eriksen}, {Fantaye}, {Farhang},
  {Fergusson}, {Fernandez-Cobos}, {Finelli}, {Forastieri}, {Frailis},
  {Fraisse}, {Franceschi}, {Frolov}, {Galeotta}, {Galli}, {Ganga},
  {G{\'e}nova-Santos}, {Gerbino}, {Ghosh}, {Gonz{\'a}lez-Nuevo}, {G{\'o}rski},
  {Gratton}, {Gruppuso}, {Gudmundsson}, {Hamann}, {Handley}, {Hansen},
  {Herranz}, {Hildebrandt}, {Hivon}, {Huang}, {Jaffe}, {Jones}, {Karakci},
  {Keih{\"a}nen}, {Keskitalo}, {Kiiveri}, {Kim}, {Kisner}, {Knox},
  {Krachmalnicoff}, {Kunz}, {Kurki-Suonio}, {Lagache}, {Lamarre}, {Lasenby},
  {Lattanzi}, {Lawrence}, {Le Jeune}, {Lemos}, {Lesgourgues}, {Levrier},
  {Lewis}, {Liguori}, {Lilje}, {Lilley}, {Lindholm}, {L{\'o}pez-Caniego},
  {Lubin}, {Ma}, {Mac{\'\i}as-P{\'e}rez}, {Maggio}, {Maino}, {Mandolesi},
  {Mangilli}, {Marcos-Caballero}, {Maris}, {Martin}, {Martinelli},
  {Mart{\'\i}nez-Gonz{\'a}lez}, {Matarrese}, {Mauri}, {McEwen}, {Meinhold},
  {Melchiorri}, {Mennella}, {Migliaccio}, {Millea}, {Mitra},
  {Miville-Desch{\^e}nes}, {Molinari}, {Montier}, {Morgante}, {Moss}, {Natoli},
  {N{\o}rgaard-Nielsen}, {Pagano}, {Paoletti}, {Partridge}, {Patanchon},
  {Peiris}, {Perrotta}, {Pettorino}, {Piacentini}, {Polastri}, {Polenta},
  {Puget}, {Rachen}, {Reinecke}, {Remazeilles}, {Renzi}, {Rocha}, {Rosset},
  {Roudier}, {Rubi{\~n}o-Mart{\'\i}n}, {Ruiz-Granados}, {Salvati}, {Sandri},
  {Savelainen}, {Scott}, {Shellard}, {Sirignano}, {Sirri}, {Spencer},
  {Sunyaev}, {Suur-Uski}, {Tauber}, {Tavagnacco}, {Tenti}, {Toffolatti},
  {Tomasi}, {Trombetti}, {Valenziano}, {Valiviita}, {Van Tent}, {Vibert},
  {Vielva}, {Villa}, {Vittorio}, {Wandelt}, {Wehus}, {White}, {White},
  {Zacchei}, \& {Zonca}}]{Planck2018}
{Planck Collaboration}, {Aghanim}, N., {Akrami}, Y., {et~al.} 2020, \aap, 641,
  A6

\bibitem[{{Rosas-Guevara} {et~al.}(2019){Rosas-Guevara}, {Bower}, {McAlpine},
  {Bonoli}, \& {Tissera}}]{Rosas-Guevara19}
{Rosas-Guevara}, Y.~M., {Bower}, R.~G., {McAlpine}, S., {Bonoli}, S., \&
  {Tissera}, P.~B. 2019, \mnras, 483, 2712

\bibitem[{{Rubinur} {et~al.}(2019){Rubinur}, {Das}, \& {Kharb}}]{Rubinur19}
{Rubinur}, K., {Das}, M., \& {Kharb}, P. 2019, \mnras, 484, 4933

\bibitem[{{Shajib} {et~al.}(2021){Shajib}, {Treu}, {Birrer}, \&
  {Sonnenfeld}}]{Shajib21}
{Shajib}, A.~J., {Treu}, T., {Birrer}, S., \& {Sonnenfeld}, A. 2021, \mnras,
  503, 2380

\bibitem[{Shen {et~al.}(2021)Shen, Chen, Hwang, Liu, Zakamska, Oguri, Li,
  Lazio, \& Breiding}]{Shen21}
Shen, Y., Chen, Y.-C., Hwang, H.-C., {et~al.} 2021, Nature Astronomy, 1,
  publisher: Nature Publishing Group

\bibitem[{{Steinborn} {et~al.}(2016){Steinborn}, {Dolag}, {Comerford},
  {Hirschmann}, {Remus}, \& {Teklu}}]{Steinborn16}
{Steinborn}, L.~K., {Dolag}, K., {Comerford}, J.~M., {et~al.} 2016, \mnras,
  458, 1013

\bibitem[{{Tozzi} {et~al.}(2021){Tozzi}, {Cresci}, {Marasco}, {Nardini},
  {Marconi}, {Mannucci}, {Chartas}, {Rizzo}, {Amiri}, {Brusa}, {Comastri},
  {Dadina}, {Lanzuisi}, {Mainieri}, {Mingozzi}, {Perna}, {Venturi}, \&
  {Vignali}}]{tozzi21}
{Tozzi}, G., {Cresci}, G., {Marasco}, A., {et~al.} 2021, \aap, 648, A99

\bibitem[{{Tremmel} {et~al.}(2017){Tremmel}, {Karcher}, {Governato},
  {Volonteri}, {Quinn}, {Pontzen}, {Anderson}, \& {Bellovary}}]{Tremmel17}
{Tremmel}, M., {Karcher}, M., {Governato}, F., {et~al.} 2017, \mnras, 470, 1121

\bibitem[{{Volonteri} {et~al.}(2003){Volonteri}, {Haardt}, \&
  {Madau}}]{Volonteri03}
{Volonteri}, M., {Haardt}, F., \& {Madau}, P. 2003, \apj, 582, 559

\bibitem[{{Volonteri} {et~al.}(2022){Volonteri}, {Pfister}, {Beckmann},
  {Dotti}, {Dubois}, {Massonneau}, {Musoke}, \& {Tremmel}}]{Volonteri22}
{Volonteri}, M., {Pfister}, H., {Beckmann}, R., {et~al.} 2022, \mnras, 514, 640

\bibitem[{{Wong}(2018)}]{wong18}
{Wong}, K.~C. 2018, in PASPC 
  Series, Vol. 514, Stellar Populations and the Distance Scale, ed.
  J.~{Jensen}, R.~M. {Rich}, \& R.~{de Grijs}, 165

\bibitem[{{Yadav} {et~al.}(2021){Yadav}, {Das}, {Barway}, \&
  {Combes}}]{Yadav21}
{Yadav}, J., {Das}, M., {Barway}, S., \& {Combes}, F. 2021, arXiv e-prints,
  arXiv:2106.12441

\end{thebibliography}


\end{document}